\documentclass[11pt]{article}
\usepackage{amsfonts}
\usepackage[final]{graphicx}

\begin{document}

\title{Zero}
\author{Gunn Quznetsov\thanks{%
mailto: gunn@mail.ru, quznets@yahoo.com}}
\date{}
\maketitle

\begin{abstract}
You shall not find any new physics, because all physical events are
interpreted well-known particles (leptons, quarks, photons, gluons,
W-bosons, Z-boson) and forces which have long known (electroweak, gravity,
strong interactions).

Contents: 1. Pointlike events and probability. 2. Leptons' moving equations
and masses. 3. Fermion-antifermion asommetry. 4. Electroweak equations. 5.
Chromatic states and gluons. 6. Asimptotic freedom, confinement, Newton's
gravity. 7. Dark energy and dark matter. 8. Events and particles. 9.
Conclusion.
\end{abstract}

\tableofcontents

\section{Pointlike events and probability}

\noindent Let $j_A:=\left\langle \mathrm{c}\rho _A,\mathbf{j}_A\right\rangle 
$ be the current 4-vector of probability density of an pointlike event $A$.

The following set of 4 complex equations with unknown $4\times 1$-matrix
complex function $\varphi (x)$ has solutions for any $A$ \cite[p.53]{PTGT}:

\[
\left\{ 
\begin{array}{c}
\varphi ^{\dagger }\varphi =\rho _A\mbox{,} \\ 
\varphi ^{\dagger }\beta ^{\left[ \nu \right] }\varphi =-\frac{j_{\nu ,A}}{%
\mathrm{c}}\mbox{.}
\end{array}
\right| 
\]
Here: $\nu \in \left\{ 1,2,3\right\} $, 
\begin{equation}
\beta ^{\left[ \nu \right] }:=\left[ 
\begin{array}{cc}
\sigma _\nu & 0_2 \\ 
0_2 & -\sigma _\nu
\end{array}
\right]  \label{bb}
\end{equation}
where $\sigma _\nu $ are the Pauli matricies, and $0_2$ is zero $2\times 2$%
-matrix.

If $\rho _A\left( x\right) =0$ for all $x$ such that $\left| x\right| \geq
\pi \mathrm{c}/\mathrm{h}$ then \cite[pp. 80--81]{Q1} $\varphi $ obeys the
following equation: 
\begin{equation}
\partial _t\varphi =c\left( \sum\limits_{\nu =1}^3\beta ^{\left[ \nu \right]
}\partial _\nu +\widehat{Q}\right) \varphi  \label{f1}
\end{equation}
where $Q_{\nu ,\mu }^{*}=-Q_{\mu ,\nu }$.

If 
\[
\gamma ^{\left[ 0\right] }:=\left[ 
\begin{array}{cc}
0_2 & 1_2 \\ 
1_2 & 0_2
\end{array}
\right] \mbox{, and }\beta ^{\left[ 4\right] }:=\mathrm{i}\left[ 
\begin{array}{cc}
0_2 & -1_2 \\ 
1_2 & 0_2
\end{array}
\right] \mbox{,} 
\]
where $1_2$ is indentical $2\times 2$-matrix, then matricies $\beta ^{\left[
\nu \right] }$ with matricies $\gamma ^{\left[ 0\right] }$ and $\beta
^{\left[ 4\right] }$ form \textit{a Clifford pentad} \cite{Q1}. There four
similar pentads exist \cite{LFTF}: I call one of them (the pentad $\beta $) 
\textit{the light} pentad, and three ($\zeta $, $\eta $, $\theta $) - 
\textit{the chromatic} pentads.

Then equation (\ref{f1}) can be rewritted as the following:

\begin{eqnarray*}
&&\rule{-0.7cm}{0pt}\biggl ( \biggr. 
\!\!{-}\!\left( \partial _0{+}\mathrm{i}\Theta _0{+}\mathrm{i}\Upsilon
_0\gamma ^{\left[ 5\right] }\right) {+}\sum\limits_{\nu =1}^3\beta ^{\left[
\nu \right] }\left( \partial _\nu {+}\mathrm{i}\Theta _\nu {+}\mathrm{i}%
\Upsilon _\nu \gamma ^{\left[ 5\right] }\right) \\
&&{+}2\left( \mathrm{i}M_0\gamma ^{\left[ 0\right] }{+}\mathrm{i}M_4\beta
^{\left[ 4\right] }\right) \biggl. \biggr ) \varphi {+} \\
&&\rule{-1cm}{0pt}{+}\biggl ( \biggr. 
\!\!{-}\!\left( \partial _0{+}\mathrm{i}\Theta _0{+}\mathrm{i}\Upsilon
_0\gamma ^{\left[ 5\right] }\right) {-}\sum\limits_{\nu =1}^3\zeta ^{[\nu
]}\left( \partial _\nu {+}\mathrm{i}\Theta _\nu {+}\mathrm{i}\Upsilon _\nu
\gamma ^{\left[ 5\right] }\right) \\
&&{+}2\left( {-}\mathrm{i}M_{\zeta ,0}\gamma _\zeta ^{[0]}{+}\mathrm{i}%
M_{\zeta ,4}\zeta ^{[4]}\right) \biggl. \biggr ) \varphi {+} \\
&&\rule{-1cm}{0pt}{+}\biggl ( \biggr.
\!\!\left( \partial _0{+}\mathrm{i}\Theta _0{+}\mathrm{i}\Upsilon _0\gamma
^{\left[ 5\right] }\right) {-}\sum\limits_{\nu =1}^3\eta ^{[\nu ]}\left(
\partial _\nu {+}\mathrm{i}\Theta _\nu {+}\mathrm{i}\Upsilon _\nu \gamma
^{\left[ 5\right] }\right) \\
&&{+}2\left( {-}\mathrm{i}M_{\eta ,0}\gamma _\eta ^{[0]}{-}\mathrm{i}M_{\eta
,4}\eta ^{[4]}\right) \biggl. \biggr ) \varphi {+} \\
&&\rule{-1cm}{0pt}{+}\biggl ( \biggr. 
\!\!{-}\!\left( \partial _0{+}\mathrm{i}\Theta _0{+}\mathrm{i}\Upsilon
_0\gamma ^{\left[ 5\right] }\right) {-}\sum\limits_{\nu =1}^3\theta ^{[\nu
]}\left( \partial _\nu {+}\mathrm{i}\Theta _\nu {+}\mathrm{i}\Upsilon _\nu
\gamma ^{\left[ 5\right] }\right) \\
&&{+}2\left( \mathrm{i}M_{\theta ,0}\gamma _\theta ^{[0]}{+}\mathrm{i}%
M_{\theta ,4}\theta ^{[4]}\right) \biggl. \biggr ) \varphi =0
\end{eqnarray*}
with real $\Theta _k\left( t,\mathbf{x}\right) $, $\Upsilon _k\left( t,%
\mathbf{x}\right) $, $M_0\left( t,\mathbf{x}\right) $, $M_4\left( t,\mathbf{x%
}\right) $, $M_{\zeta ,0}\left( t,\mathbf{x}\right) $, $M_{\zeta ,4}\left( t,%
\mathbf{x}\right) $,\linebreak $M_{\eta ,0}\left( t,\mathbf{x}\right) $, $%
M_{\eta ,4}\left( t,\mathbf{x}\right) $, $M_{\theta ,0}\left( t,\mathbf{x}%
\right) $, $M_{\theta ,4}\left( t,\mathbf{x}\right) $ and with 
\[
\gamma ^{\left[ 5\right] }:=\left[ 
\begin{array}{cc}
1_2 & 0_2 \\ 
0_2 & -1_2
\end{array}
\right] \mbox{.} 
\]

First summand of this equation contains elements of the light pentad, only.
And the rest summands contain elements of the chromatic pentads, only.

This equation can be rewritten as the following: 
\begin{eqnarray}
&&\Bigl(-\left( \partial _0+\mathrm{i}\Theta _0+\mathrm{i}\Upsilon _0\gamma
^{\left[ 5\right] }\right) +\sum\limits_{\nu =1}^3\beta ^{\left[ \nu \right]
}\left( \partial _\nu +\mathrm{i}\Theta _\nu +\mathrm{i}\Upsilon _\nu \gamma
^{\left[ 5\right] }\right) +  \label{ham0} \\
&&+\mathrm{i}M_0\gamma ^{\left[ 0\right] }+\mathrm{i}M_4\beta ^{\left[
4\right] }-  \nonumber \\
&&-\mathrm{i}M_{\zeta ,0}\gamma _\zeta ^{[0]}+\mathrm{i}M_{\zeta ,4}\zeta
^{[4]}  \nonumber \\
&&-\mathrm{i}M_{\eta ,0}\gamma _\eta ^{[0]}-\mathrm{i}M_{\eta ,4}\eta ^{[4]}
\nonumber \\
&&+\mathrm{i}M_{\theta ,0}\gamma _\theta ^{[0]}+\mathrm{i}M_{\theta
,4}\theta ^{[4]}\Bigr)\varphi =0  \nonumber
\end{eqnarray}
because 
\[
\zeta ^{[\nu ]}{+}\eta ^{[\nu ]}{+}\theta ^{[\nu ]}={-}\beta ^{\left[ \nu
\right] }\mbox{.} 
\]

This equatin is a generalization of the Dirac's equation with gauge fields $%
\Theta _k\left( t,\mathbf{x}\right) $ and $\Upsilon _k\left( t,\mathbf{x}%
\right) $ and with eight mass members. The mass members with elements of the
light pentad ( $M_0 $ and $M_4$) conform to neutrino and its' lepton states 
\cite{n}. And six mass members with elements of chromatic pentads conform to
three pairs (up and down) of chromatic states (red, green, blue).

\section{Leptons' moving equation and masses}

I call the following part of equation (\ref{ham0}) 
\begin{eqnarray}
&&\Bigl(-\left( \partial _0+\mathrm{i}\Theta _0+\mathrm{i}\Upsilon _0\gamma
^{\left[ 5\right] }\right) +\sum\limits_{\nu =1}^3\beta ^{\left[ \nu \right]
}\left( \partial _\nu +\mathrm{i}\Theta _\nu +\mathrm{i}\Upsilon _\nu \gamma
^{\left[ 5\right] }\right)  \label{ham1} \\
&&+\mathrm{i}M_0\gamma ^{\left[ 0\right] }+\mathrm{i}M_4\beta ^{\left[
4\right] }\Bigr)\varphi =0  \nonumber
\end{eqnarray}
\textit{a lepton moving equation}.

Let 
\[
u_{A,\nu }:=\mathrm{c}\frac{j_{A,\nu }}{\rho _A}\mbox{, }u_{A,4}:=\mathrm{c}%
\frac{\varphi ^{\dagger }\beta ^{\left[ 4\right] }\varphi }{\rho _A}\mbox{, }%
u_{A,5}:=\mathrm{c}\frac{\varphi ^{\dagger }\gamma ^{\left[ 0\right]
}\varphi }{\rho _A}\mbox{. } 
\]

In this case 
\[
\sum\limits_{k=1}^5u_{A,k}^2=\mathrm{c}^2\mbox{.} 
\]

Thus, only all five elements of a Clifford pentad provide an entire set of
speed components and, for completeness, yet two ''space'' coordinates $x_4$
and $x_5$ should be added to our three $x_1$, $x_2$, $x_3$. These additional
coordinates can be selected so that 
\[
-\frac{\pi \mathrm{c}}{\mathrm{h}}\leq x_4\leq \frac{\pi \mathrm{c}}{\mathrm{%
h}}\mbox{, }-\frac{\pi \mathrm{c}}{\mathrm{h}}\leq x_5\leq \frac{\pi \mathrm{%
c}}{\mathrm{h}}\mbox{.} 
\]

Coordinates $x_4$ and $x_5$ are not coordinates of any events. Hence, our
devices do not detect them as actual space coordinates.

Denote: 
\begin{eqnarray*}
&&\widetilde{\varphi }\left( x_1,x_2,x_3,x_4,x_5\right) := \\
&&:=\widetilde{\varphi }\left( x_1,x_2,x_3\right) \exp \left( \mathrm{i}%
\left( x_5M_0\left( x_1,x_2,x_3\right) +x_4M_4\left( x_1,x_2,x_3\right)
\right) \right) \mbox{.}
\end{eqnarray*}

In that case equation (\ref{ham1}) has the following shape: 
\begin{eqnarray}
&&\Bigl(-\left( \mathrm{i}\partial _0+\Theta _0+\Upsilon _0\gamma ^{\left[
5\right] }\right) +\sum\limits_{\nu =1}^3\beta ^{\left[ \nu \right] }\left( 
\mathrm{i}\partial _\nu +\Theta _\nu +\Upsilon _\nu \gamma ^{\left[ 5\right]
}\right)  \label{ham2} \\
&&+\left( -\gamma ^{\left[ 0\right] }\mathrm{i}\partial _5-\beta ^{\left[
4\right] }\mathrm{i}\partial _4\right) \Bigr)\widetilde{\varphi } =0 \mbox{.}
\nonumber
\end{eqnarray}

Let $\beta ^{\left[ 0\right] }:=-1_4$ .

Because the following system of equations with unknown functions $B_k$ and $%
F_k$ 
\[
\left\{ 
\begin{array}{c}
0.5g_1B_k-F_k=-\Theta _k-\Upsilon _k\mbox{,} \\ 
g_1B_k-F_k=-\Theta _k+\Upsilon _k
\end{array}
\right\| 
\]
has solution for any $k$ and for some constant real positive number $g_1$
then equation (\ref{ham2}) has got the following form: 
\begin{equation}
\sum\limits_{k=0}^3\beta ^{\left[ k\right] }\left( \mathrm{i}\partial
_k+F_k+0.5g_1YB_k\right) \widetilde{\varphi }-\left( \gamma ^{\left[
0\right] }\mathrm{i}\partial _5+\beta ^{\left[ 4\right] }\mathrm{i}\partial
_4\right) )\widetilde{\varphi }=0  \label{ham3}
\end{equation}
with 
\[
Y:=-\left[ 
\begin{array}{cc}
1_2 & 0_2 \\ 
0_2 & 2\cdot 1_2
\end{array}
\right] \mbox{.} 
\]

Let 
\[
N_\vartheta :=\mathrm{trunk}\left( \frac{\mathrm{c}}{\mathrm{h}}M_0\right) %
\mbox{, }N_\omega :=\mathrm{trunk}\left( \frac{\mathrm{c}}{\mathrm{h}}%
M_4\right) \mbox{.} 
\]

In this case Fourier series for $\widetilde{\varphi }$ is of the following
form: 
\begin{equation}
\widetilde{\varphi }\left( t,\mathbf{x},x_4,x_5\right) =\varphi \left( t,%
\mathbf{x}\right) \sum_{n,s}\delta _{-n,N_\vartheta \left( t,\mathbf{x}%
\right) }\delta _{-s,N_\varpi \left( t,\mathbf{x}\right) }\exp \left( -%
\mathrm{i}\frac{\mathrm{h}}{\mathrm{c}}\left( nx_5+sx_4\right) \right) %
\mbox{.}  \label{ff}
\end{equation}

From properties of $\delta $: in every point $\left\langle t,\mathbf{x}%
\right\rangle $ : either $\widetilde{\varphi }\left( t,\mathbf{x}%
,x_4,x_5\right) $=0 or integer numbers $n_0$ and $s_0$ exist for which: 
\[
\widetilde{\varphi }\left( t,\mathbf{x},x_4,x_5\right) =\varphi \left( t,%
\mathbf{x}\right) \exp \left( -\mathrm{i}\frac{\mathrm{h}}{\mathrm{c}}\left(
n_0x_5+s_0x_4\right) \right) \mbox{.} 
\]

That is for every space-time point: either this point is empty or single
mass is placed in this point.

Let $N_\varpi \left( t,\mathbf{x}\right) =0$ and $N_\vartheta \left( t,%
\mathbf{x}\right) =n_0$. In that case: 
\[
\widetilde{\varphi }\left( t,\mathbf{x},x_4,x_5\right) =\varphi \left( t,%
\mathbf{x}\right) \exp \left( -\mathrm{i}\frac{\mathrm{h}}{\mathrm{c}}%
n_0x_5\right) \mbox{.} 
\]

\section{Fermion-antifermion asimmetry}

Let $\Im _0$ be a space spanned of sub basis with the following elements: 
\[
\frac{\mathrm{h}}{2\pi \mathrm{c}}\exp \left( -\mathrm{i}\frac{\mathrm{h}}{%
\mathrm{c}}n_0x_5\right) \varepsilon _k 
\]
(here $\varepsilon _k$ are vectors of the basis in which matricies $\beta
^{\left[ \nu \right] }$ have form (\ref{bb})).

In this case if 
\[
\widehat{H}_0:=\mathrm{c}\sum\limits_{\nu =1}^3\beta ^{\left[ \nu \right] }%
\mathrm{i}\partial _\nu +\mathrm{h}n_0\gamma ^{\left[ 0\right] }
\]
then $\widehat{H}_0$ is the free Hamiltonian on $\Im _0$.

Let $\mathbf{k}$ be a vector $\left\langle k_1,k_2,k_3\right\rangle $ where $%
k_s$ are integer numbers and let 
\[
\omega \left( \mathbf{k}\right) :=\sqrt{k_1^2+k_2^2+k_3^2+n^2} 
\]
where $n$ is a natural number.

Let 
\[
e_1\left( \mathbf{k}\right) :=\frac 1{2\sqrt{\omega \left( \mathbf{k}\right)
\left( \omega \left( \mathbf{k}\right) +n\right) }}\left[ 
\begin{array}{c}
\omega \left( \mathbf{k}\right) +n+k_3 \\ 
k_1+\mathrm{i}k_2 \\ 
\omega \left( \mathbf{k}\right) +n-k_3 \\ 
-k_1-\mathrm{i}k_2
\end{array}
\right] \mbox{,} 
\]

\[
e_2\left( \mathbf{k}\right) :=\frac 1{2\sqrt{\omega \left( \mathbf{k}\right)
\left( \omega \left( \mathbf{k}\right) +n\right) }}\left[ 
\begin{array}{c}
k_1-\mathrm{i}k_2 \\ 
\omega \left( \mathbf{k}\right) +n-k_3 \\ 
-k_1-\mathrm{i}k_2 \\ 
\omega \left( \mathbf{k}\right) +n+k_3
\end{array}
\right] \mbox{,} 
\]

\[
e_3\left( \mathbf{k}\right) :=\frac 1{2\sqrt{\omega \left( \mathbf{k}\right)
\left( \omega \left( \mathbf{k}\right) +n\right) }}\left[ 
\begin{array}{c}
-\omega \left( \mathbf{k}\right) -n+k_3 \\ 
k_1+\mathrm{i}k_2 \\ 
\omega \left( \mathbf{k}\right) +n+k_3 \\ 
k_1+\mathrm{i}k_2
\end{array}
\right] \mbox{,} 
\]

\[
e_4\left( \mathbf{k}\right) :=\frac 1{2\sqrt{\omega \left( \mathbf{k}\right)
\left( \omega \left( \mathbf{k}\right) +n\right) }}\left[ 
\begin{array}{c}
k_1-\mathrm{i}k_2 \\ 
-\omega \left( \mathbf{k}\right) -n-k_3 \\ 
k_1-\mathrm{i}k_2 \\ 
\omega \left( \mathbf{k}\right) +n-k_3
\end{array}
\right] \mbox{.} 
\]

In that case functions 
\[
e_1(\mathbf{k})(2\mathrm{c}/\mathrm{h})^{3/2}\exp (-\mathrm{i}(\mathrm{h}/%
\mathrm{c})\mathbf{kx}) \mbox{ and } e_2(\mathbf{k})(2\mathrm{c}/\mathrm{h}%
)^{3/2}\exp (-\mathrm{i}(\mathrm{h}/\mathrm{c})\mathbf{kx}) 
\]
are eigenvectors of $\widehat{H}_0$ with eigenvalues $(+\mathrm{h}\omega (%
\mathbf{k}))$, and functions 
\[
e_3(\mathbf{k})(2\mathrm{c}/\mathrm{h})^{3/2}\exp (-\mathrm{i}(\mathrm{h}/%
\mathrm{c})\mathbf{kx}) \mbox{ and } e_4(\mathbf{k})(2\mathrm{c}/\mathrm{h}%
)^{3/2}\exp (-\mathrm{i}(\mathrm{h}/\mathrm{c})\mathbf{kx}) 
\]
are eigenvectors of $\widehat{H}_0$with eigenvalues $(-\mathrm{h}\omega (%
\mathbf{k}))$.

Let $\frak{H}$ be some unitary space. Let $\widetilde{0}$ be the zero
element of $\frak{H}$. That is any element $\widetilde{F}$ of $\frak{H}$
obeys to the following conditions:

\vspace*{4mm} $0\widetilde{F}=\widetilde{0}$, $\widetilde{0}+\widetilde{F}=%
\widetilde{F}$, $\widetilde{0}^{\dagger }\widetilde{F}=\widetilde{F}$, $%
\widetilde{0}^{\dagger }=\widetilde{0}$.

\vspace*{4mm} Let $\widehat{0}$ be the zero operator on $\frak{H}$. That is
any element $\widetilde{F}$ of $\frak{H}$ obeys to the following condition:

\vspace*{4mm} $\widehat{0}\widetilde{F}=0\widetilde{F}$, and if $\widehat{b}$
is any operator on $\frak{H}$ then

\vspace*{4mm} $\widehat{0}+\widehat{b}=\widehat{b}+\widehat{0}=\widehat{b}$, 
$\widehat{0}\widehat{b}=\widehat{b}\widehat{0}=\widehat{0}$.

\vspace*{4mm} Let $\widehat{1}$ be the identy operator on $\frak{H}$. That
is any element $\widetilde{F}$ of $\frak{H}$ obeys to the following
condition:

$\widehat{1}\widetilde{F}=1\widetilde{F}=\widetilde{F}$ , and if $\widehat{b}
$ is any operator on $\frak{H}$ then\\ $\widehat{1}\widehat{b}=\widehat{b}%
\widehat{1}=$ $\widehat{b}$.

Let linear operators $b_{s,\mathbf{k}}$ ($s\in \left\{ 1,2,3,4\right\} $)
act on all elements of this space. And let these operators fulfill the
following conditions: 
\[
\left\{ b_{s,\mathbf{k}}^{\dagger },b_{s^{\prime },\mathbf{k}^{\prime
}}\right\} :=b_{s,\mathbf{k}}^{\dagger }b_{s^{\prime },\mathbf{k}^{\prime
}}+b_{s^{\prime },\mathbf{k}^{\prime }}b_{s,\mathbf{k}}^{\dagger }=\left( 
\frac{\mathrm{h}}{2\pi {}}\right) ^3\delta _{\mathbf{k},\mathbf{k}^{\prime
}}\delta _{s,s^{\prime }}\widehat{1}\mbox{,}
\]
\[
\left\{ b_{s,\mathbf{k}},b_{s^{\prime },\mathbf{k}^{\prime }}\right\} =b_{s,%
\mathbf{k}}b_{s^{\prime },\mathbf{k}^{\prime }}+b_{s^{\prime },\mathbf{k}%
^{\prime }}b_{s,\mathbf{k}}=\left\{ b_{s,\mathbf{k}}^{\dagger },b_{s^{\prime
},\mathbf{k}^{\prime }}^{\dagger }\right\} =\widehat{0}\mbox{.}
\]

Hence, 
\[
b_{s,\mathbf{k}}b_{s,\mathbf{k}}=b_{s,\mathbf{k}}^{\dagger }b_{s,\mathbf{k}%
}^{\dagger }=\widehat{0}\mbox{.} 
\]

There exists element $\widetilde{F}_0$ of $\frak{H}$ such that $\widetilde{F}%
_0^{\dagger }\widetilde{F}_0=1$ and for any $b_{s,\mathbf{k}}$: $b_{s,%
\mathbf{k}}\widetilde{F}_0=\widetilde{0}$. Hence, $\widetilde{F}_0^{\dagger
}b_{s,\mathbf{k}}^{\dagger }=\widetilde{0}$.

Let 
\[
\psi _s\left( \mathbf{x}\right) :=\sum_{\mathbf{k}}\sum\limits_{r=1}^4b_{r,%
\mathbf{k}}e_{r,s}\left( \mathbf{k}\right) \exp \left( -\mathrm{i}\frac{%
\mathrm{h}}{\mathrm{c}}\mathbf{kx}\right) \mbox{.} 
\]

And these operators obey the following conditions:

\begin{eqnarray*}
\left\{ \psi _s^{\dagger }\left( \mathbf{x}\right) ,\psi _{s^{\prime
}}\left( \mathbf{x}^{\prime }\right) \right\}&:=&\psi _s^{\dagger }\left( 
\mathbf{x}\right) \psi _{s^{\prime }}\left( \mathbf{x}^{\prime }\right)
+\psi _{s^{\prime }}\left( \mathbf{x}^{\prime }\right) \psi _s^{\dagger
}\left( \mathbf{x}\right) \\
&=&\delta \left( \mathbf{x-x}^{\prime }\right) \delta _{s,s^{\prime }}%
\widehat{1}\mbox{.}
\end{eqnarray*}

\vspace*{4mm} $\psi _s\left( \mathbf{x}\right) \widetilde{F}_0=\widetilde{0}$%
, $\left\{ \psi _s\left( \mathbf{x}\right) ,\psi _{s^{\prime }}\left( 
\mathbf{x}^{\prime }\right) \right\} =\left\{ \psi _s^{\dagger }\left( 
\mathbf{x}\right) ,\psi _{s^{\prime }}^{\dagger }\left( \mathbf{x}^{\prime
}\right) \right\} =\widehat{0}$.

\vspace*{4mm} Hence,

\vspace*{4mm} $\psi _s\left( \mathbf{x}\right) \psi _{s^{\prime }}\left( 
\mathbf{x}^{\prime }\right) =\psi _s^{\dagger }\left( \mathbf{x}\right) \psi
_{s^{\prime }}^{\dagger }\left( \mathbf{x}^{\prime }\right) =\widehat{0}$.

\vspace*{4mm} Let 
\[
\Psi \left( t,\mathbf{x}\right) :=\sum\limits_{s=1}^4\varphi _s\left( t,%
\mathbf{x}\right) \psi _s^{\dagger }\left( \mathbf{x}\right) \widetilde{F}_0%
\mbox{.} 
\]

These function obey the following condition: 
\[
\Psi ^{\dagger }\left( t,\mathbf{x}^{\prime }\right) \Psi \left( t,\mathbf{x}%
\right) =\varphi ^{\dagger }\left( t,\mathbf{x}^{\prime }\right) \varphi
\left( t,\mathbf{x}\right) \delta \left( \mathbf{x-x}^{\prime }\right) %
\mbox{.} 
\]

Hence, 
\[
\int d\mathbf{x}^{\prime }\cdot \Psi ^{\dagger }\left( t,\mathbf{x}^{\prime
}\right) \Psi \left( t,\mathbf{x}\right) =\rho \left( t,\mathbf{x}\right) %
\mbox{.} 
\]

Let a Fourier series of $\varphi _s\left( t,\mathbf{x}\right) $ has the
following form: 
\[
\varphi _s\left( t,\mathbf{x}\right) =\sum_{\mathbf{p}}\sum%
\limits_{r=1}^4c_r\left( t,\mathbf{p}\right) e_{r,s}\left( \mathbf{p}\right)
\exp \left( -\mathrm{i}\frac{\mathrm{h}}{\mathrm{c}}\mathbf{px}\right) %
\mbox{.} 
\]

In that case: 
\[
\underline{\Psi }\left( t,\mathbf{p}\right) :=\left( \frac{2\pi \mathrm{c}}{%
\mathrm{h}}\right) ^3\sum\limits_{r=1}^4c_r\left( t,\mathbf{p}\right) b_{r,%
\mathbf{p}}^{\dagger }\widetilde{F}_0\mbox{.} 
\]

If 
\begin{equation}
\mathcal{H}_0\left( \mathbf{x}\right) :=\psi ^{\dagger }\left( \mathbf{x}%
\right) \widehat{H}_0\psi \left( \mathbf{x}\right)   \label{h}
\end{equation}
then $\mathcal{H}_0\left( \mathbf{x}\right) $ is called a Hamiltonian $%
\widehat{H}_0$ density.

Because 
\[
\widehat{H}_0\varphi \left( t,\mathbf{x}\right) =\mathrm{i}\frac \partial
{\partial t}\varphi \left( t,\mathbf{x}\right) 
\]
then 
\begin{equation}
\int d\mathbf{x}^{\prime }\cdot \mathcal{H}_0\left( \mathbf{x}^{\prime
}\right) \Psi \left( t,\mathbf{x}\right) =\mathrm{i}\frac \partial {\partial
t}\Psi \left( t,\mathbf{x}\right) \mbox{.}  \label{ham}
\end{equation}

Therefore, if 
\[
\widehat{\Bbb{H}}:=\int d\mathbf{x}^{\prime }\cdot \mathcal{H}_0\left( 
\mathbf{x}^{\prime }\right) 
\]

then $\widehat{\Bbb{H}}$ acts similar to the Hamiltonian on space $\frak{H}$.

And if 
\[
E_\Psi \left( \widetilde{F}_0\right) :=\sum_{\mathbf{p}}\underline{\Psi }%
^{\dagger }\left( t,\mathbf{p}\right) \widehat{\Bbb{H}}\underline{\Psi }%
\left( t,\mathbf{p}\right) 
\]
then $E_\Psi \left( \widetilde{F}_0\right) $ is an energy of $\Psi $ on
vacuum $\widetilde{F}_0$.

Operator $\widehat{\Bbb{H}}$ obeys the following condition: 
\[
\widehat{\Bbb{H}}=\left( \frac{2\pi \mathrm{c}}{\mathrm{h}}\right) ^3\sum_{%
\mathbf{k}}\mathrm{h}\omega \left( \mathbf{k}\right) \left(
\sum\limits_{r=1}^2b_{r,\mathbf{k}}^{\dagger }b_{r,\mathbf{k}%
}-\sum\limits_{r=3}^4b_{r,\mathbf{k}}^{\dagger }b_{r,\mathbf{k}}\right) %
\mbox{.}
\]

This operator is not positive defined and in this case 
\[
E_\Psi \left( \widetilde{F}_0\right) =\left( \frac{2\pi \mathrm{c}}{\mathrm{h%
}}\right) ^3\sum_{\mathbf{p}}\mathrm{h}\omega \left( \mathbf{p}\right)
\left( \sum\limits_{r=1}^2\left| c_r\left( t,\mathbf{p}\right) \right|
^2-\sum\limits_{r=3}^4\left| c_r\left( t,\mathbf{p}\right) \right| ^2\right) %
\mbox{.} 
\]

This problem is usually solved in the following way \cite[p.54]{psk}:

Let: 
\begin{eqnarray*}
v_1\left( \mathbf{k}\right) &:&=\gamma ^{\left[ 0\right] }e_3\left( \mathbf{k%
}\right) \mbox{,} \\
v_2\left( \mathbf{k}\right) &:&=\gamma ^{\left[ 0\right] }e_4\left( \mathbf{k%
}\right) \mbox{,} \\
d_{1,\mathbf{k}} &:&=-b_{3,-\mathbf{k}}^{\dagger }\mbox{,} \\
d_{2,\mathbf{k}} &:&=-b_{4,-\mathbf{k}}^{\dagger }\mbox{.}
\end{eqnarray*}

Therefore, 
\begin{eqnarray*}
\psi _s\left( \mathbf{x}\right)  &:&=\sum_{\mathbf{k}}\sum\limits_{r=1}^2%
\Bigl(b_{r,\mathbf{k}}e_{r,s}\left( \mathbf{k}\right) \exp \left( -\mathrm{i}%
\frac{\mathrm{h}}{\mathrm{c}}\mathbf{kx}\right) + \\
&&+d_{r,\mathbf{k}}^{\dagger }v_{r,s}\left( \mathbf{k}\right) \exp \left( 
\mathrm{i}\frac{\mathrm{h}}{\mathrm{c}}\mathbf{kx}\right) \Bigr)
\end{eqnarray*}
\begin{eqnarray*}
\widehat{\Bbb{H}} &=&\left( \frac{2\pi \mathrm{c}}{\mathrm{h}}\right)
^3\sum_{\mathbf{k}}\mathrm{h}\omega \left( \mathbf{k}\right)
\sum\limits_{r=1}^2\left( b_{r,\mathbf{k}}^{\dagger }b_{r,\mathbf{k}}+d_{r,%
\mathbf{k}}^{\dagger }d_{r,\mathbf{k}}\right)  \\
&&-2\sum_{\mathbf{k}}\mathrm{h}\omega \left( \mathbf{k}\right) \widehat{1}%
\mbox{.}
\end{eqnarray*}

The first term on the right side of this equality is positive defined. This
term is taken as the desired Hamiltonian. The second term of this equality
is infinity constant. And this infinity is deleted (?!) \cite[p.58]{psk}

But in this case $d_{r,\mathbf{k}}\widetilde{F}_0\neq \widetilde{0}$. In
order to satisfy such condition, the vacuum element $\widetilde{F}_0$ must
be replaced by the following: 
\[
\widetilde{F}_0\rightarrow \widetilde{\Phi }_0:=\prod_{\mathbf{k}%
}\prod\limits_{r=3}^4\left( \frac{2\pi \mathrm{c}}{\mathrm{h}}\right) ^3b_{r,%
\mathbf{k}}^{\dagger }\widetilde{F}_0\mbox{.} 
\]

But in this case: 
\[
\psi _s\left( \mathbf{x}\right) \widetilde{\Phi }_0\neq \widetilde{0}\mbox{.}
\]
And condition (\ref{ham}) isn't carried out.

In order to satisfy such condition, operators $\psi _s\left( \mathbf{x}%
\right) $ must be replaced by the following: 
\begin{eqnarray*}
\rule{-1,0cm}{0pt}&&\psi _s\left( \mathbf{x}\right) \rightarrow \phi
_s\left( \mathbf{x}\right) \!:=\! \\
\rule{-1,0cm}{0pt}&&\!:=\!\sum_{\mathbf{k}}\sum\limits_{r=1}^2\Bigl(b_{r,%
\mathbf{k}}e_{r,s}\left( \mathbf{k}\right) \exp \left( -\mathrm{i}\frac{%
\mathrm{h}}{\mathrm{c}}\mathbf{kx}\right) +d_{r,\mathbf{k}}v_r\left( \mathbf{%
k}\right) \exp \left( \mathrm{i}\frac{\mathrm{h}}{\mathrm{c}}\mathbf{kx}%
\right) \Bigr)\mbox{.}
\end{eqnarray*}

Hence, 
\begin{eqnarray*}
\widehat{\Bbb{H}} &=&\int d\mathbf{x}\cdot \mathcal{H}\left( \mathbf{x}%
\right) =\int d\mathbf{x}\cdot \phi ^{\dagger }\left( \mathbf{x}\right) 
\widehat{H}_0\phi \left( \mathbf{x}\right) = \\
&=&\left( \frac{2\pi \mathrm{c}}{\mathrm{h}}\right) ^3\sum_{\mathbf{k}}%
\mathrm{h}\omega \left( \mathbf{k}\right) \sum\limits_{r=1}^2\left( b_{r,%
\mathbf{k}}^{\dagger }b_{r,\mathbf{k}}-d_{r,\mathbf{k}}^{\dagger }d_{r,%
\mathbf{k}}\right) \mbox{.}
\end{eqnarray*}

And again we get negative energy.

Let's consider the meaning of such energy: An event with positive energy
transfers this energy photons which carries it on recorders observers.
Observers know that this event occurs, not before it happens. But event with
negative energy should absorb this energy from observers. Consequently,
observers know that this event happens before it happens. This contradicts
Theorem 3.4.2 \cite[pp.34--35]{LFTF}. Therefore, events with negative energy
do not occur.

Hence, over vacuum $\widetilde{\Phi }_0$ single fermions can exist, but
there is no single antifermions.

A two-particle state is defined the following field operator \cite{Z}: 
\[
\psi _{s_1,s_2}\left( \mathbf{x,y}\right) :=\left| 
\begin{array}{cc}
\phi _{s_1}\left( \mathbf{x}\right) & \phi _{s_2}\left( \mathbf{x}\right) \\ 
\phi _{s_1}\left( \mathbf{y}\right) & \phi _{s_2}\left( \mathbf{y}\right)
\end{array}
\right| \mbox{.} 
\]

In that case: 
\[
\widehat{\Bbb{H}}=2\mathrm{h}\left( \frac{2\pi \mathrm{c}}{\mathrm{h}}%
\right) ^6\left( \widehat{\Bbb{H}}_a+\widehat{\Bbb{H}}_b\right) 
\]
where 
\begin{eqnarray*}
\widehat{\Bbb{H}}_a &:&=\sum_{}\sum_{}\left( \omega \left( \mathbf{k}\right)
-\omega \left( \mathbf{p}\right) \right) \sum_{r=1}^2\sum_{j=1}^2\times  \\
&&\ \times \Bigl\{v_j^{\dagger }\left( -\mathbf{k}\right) v_j\left( -\mathbf{%
p}\right) e_r^{\dagger }\left( \mathbf{p}\right) e_r\left( \mathbf{k}\right)
\times  \\
&&\ \times \left( +b_{r,\mathbf{p}}^{\dagger }d_{j,-\mathbf{k}}^{\dagger
}d_{j,-\mathbf{p}}b_{r,\mathbf{k}}\right) + \\
&&\ +\left( +d_{r,-\mathbf{p}}^{\dagger }b_{j,\mathbf{k}}^{\dagger }b_{j,%
\mathbf{k}}d_{r,-\mathbf{p}}\right) + \\
&&\ +v_j^{\dagger }\left( -\mathbf{p}\right) v_j\left( -\mathbf{k}\right)
e_r^{\dagger }\left( \mathbf{k}\right) e_r\left( \mathbf{p}\right) \times  \\
&&\ \times \left( -b_{r,\mathbf{k}}^{\dagger }d_{j,-\mathbf{p}}^{\dagger
}d_{j,-\mathbf{k}}b_{r,\mathbf{p}}\right) + \\
&&\ +\left( -b_{r,\mathbf{p}}^{\dagger }d_{j,-\mathbf{k}}^{\dagger }d_{j,-%
\mathbf{k}}b_{r,\mathbf{p}}\right) \Bigr\}
\end{eqnarray*}
and 
\begin{eqnarray*}
\widehat{\Bbb{H}}_b &:&=\sum_{}\sum_{}\left( \omega \left( \mathbf{k}\right)
+\omega \left( \mathbf{p}\right) \right) \sum_{r=1}^2\sum_{j=1}^2\times  \\
&&\ \times \Bigl\{v_j^{\dagger }\left( -\mathbf{p}\right) v_j\left( -\mathbf{%
k}\right) v_r^{\dagger }\left( -\mathbf{k}\right) v_r\left( -\mathbf{p}%
\right) \times  \\
&&\ \times \left( -d_{r,-\mathbf{k}}^{\dagger }d_{j,-\mathbf{p}}^{\dagger
}d_{j,-\mathbf{k}}d_{r,-\mathbf{p}}\right) + \\
&&\ +\left( -d_{r,-\mathbf{p}}^{\dagger }d_{j,-\mathbf{k}}^{\dagger }d_{j,-%
\mathbf{k}}d_{r,-\mathbf{p}}\right)  \\
&&\ +e_r^{\dagger }\left( \mathbf{k}\right) e_r\left( \mathbf{p}\right)
e_j^{\dagger }\left( \mathbf{p}\right) e_j\left( \mathbf{k}\right) \times  \\
&&\ \times \left( +b_{r,\mathbf{k}}^{\dagger }b_{j,\mathbf{p}}^{\dagger
}b_{j,\mathbf{k}}b_{r,\mathbf{p}}\right) + \\
&&\ +\left( +b_{r,\mathbf{p}}^{\dagger }b_{j,\mathbf{k}}^{\dagger }b_{j,%
\mathbf{k}}b_{r,\mathbf{p}}\right) \Bigr\}\mbox{.}
\end{eqnarray*}

If velocities are small then the following formula is fair. 
\[
\widehat{\Bbb{H}}=4\mathrm{h}\left( \frac{2\pi \mathrm{c}}{\mathrm{h}}%
\right) ^6\left( \widehat{\Bbb{H}}_a+\widehat{\Bbb{H}}_b\right) 
\]
where 
\begin{eqnarray*}
\widehat{\Bbb{H}}_a &:&=\sum_{\mathbf{k}}\sum_{\mathbf{p}}\left( \omega
\left( \mathbf{k}\right) -\omega \left( \mathbf{p}\right) \right) \times  \\
&&\times \sum_{r=1}^2\sum_{j=1}^2\left( d_{j,-\mathbf{p}}^{\dagger }b_{r,%
\mathbf{k}}^{\dagger }b_{r,\mathbf{k}}d_{j,-\mathbf{p}}-b_{j,\mathbf{p}%
}^{\dagger }d_{r,-\mathbf{k}}^{\dagger }d_{r,-\mathbf{k}}b_{j,\mathbf{p}%
}\right) 
\end{eqnarray*}
and 
\begin{eqnarray*}
\widehat{\Bbb{H}}_b &:&=\sum_{\mathbf{k}}\sum_{\mathbf{p}}\left( \omega
\left( \mathbf{k}\right) +\omega \left( \mathbf{p}\right) \right) \times  \\
&&\times \sum_{j=1}^2\sum_{r=1}^2\left( b_{j,\mathbf{p}}^{\dagger }b_{r,%
\mathbf{k}}^{\dagger }b_{r,\mathbf{k}}b_{j,\mathbf{p}}-d_{j,-\mathbf{p}%
}^{\dagger }d_{r,-\mathbf{k}}^{\dagger }d_{r,-\mathbf{k}}d_{j,-\mathbf{p}%
}\right) \mbox{.}
\end{eqnarray*}

Therefore, in any case events with pairs of fermions and events with
fermion-antifermion pairs can occur, but events with pairs of antiftrmions
can not happen.

\section{Electroweak equations}

Let us consider the space $\Im _{e,\nu }$ spanned of the following basis 
\cite[p.18]{NH}: 
\[
{}_{e,\nu }:=\left\{ 
\begin{array}{c}
\frac{\mathrm{h}}{2\pi \mathrm{c}}\sqrt{\frac{2\pi n_0}{\sinh \left( 2\pi
n_0\right) }}\left( \cosh \left( \frac{\mathrm{h}}{\mathrm{c}}n_0x_4\right)
+\sinh \left( \frac{\mathrm{h}}{\mathrm{c}}n_0x_4\right) \right) \varepsilon
_k, \\ 
\frac{\mathrm{h}}{2\pi \mathrm{c}}\exp \left( -i\frac{\mathrm{h}}{\mathrm{c}}%
n_0x_5\right) \varepsilon _k
\end{array}
\right\} 
\]
with some integer $n_0$.

In this space equation (\ref{ham3}) is equivalent to the following 
\cite[p.19]{NH}: 
\[
\sum\limits_{k=0}^3\beta ^{\left[ k\right] }\left( \mathrm{i}\partial _k+e%
\widehat{A}_k+0.5\left( \widehat{Z}_k+\widehat{W}_k\right) \right) 
\widetilde{\varphi }-\left( \gamma ^{\left[ 0\right] }\mathrm{i}\partial
_5+\beta ^{\left[ 4\right] }\mathrm{i}\partial _4\right) \widetilde{\varphi }%
=0 
\]
with 
\[
\widetilde{\varphi }:=\left[ 
\begin{array}{c}
\varphi _{\nu ,1} \\ 
\varphi _{\nu ,2} \\ 
0 \\ 
0 \\ 
\varphi _{e,L,1} \\ 
\varphi _{e,L,2} \\ 
\varphi _{e,R,1} \\ 
\varphi _{e,R,2}
\end{array}
\right] \mbox{, }\widehat{A}_k:=A_k\left[ 
\begin{array}{cccc}
0_2 & 0_2 & 0_2 & 0_2 \\ 
0_2 & 1_2 & 0_2 & 0_2 \\ 
0_2 & 0_2 & 1_2 & 0_2 \\ 
0_2 & 0_2 & 0_2 & 1_2
\end{array}
\right] \mbox{,} 
\]
\[
\widehat{Z}_k:=Z_k\frac 1{\sqrt{g_1^2+g_2^2}}\left[ 
\begin{array}{cccc}
\left( g_1^2+g_2^2\right) 1_2 & 0_2 & 0_2 & 0_2 \\ 
0_2 & 2g_1^21_2 & 0_2 & 0_2 \\ 
0_2 & 0_2 & \left( g_2^2-g_1^2\right) 1_2 & 0_2 \\ 
0_2 & 0_2 & 0_2 & 2g_1^21_2
\end{array}
\right] \mbox{,} 
\]
\[
\widehat{W}_k:=g_2\left[ 
\begin{array}{cccc}
0_2 & 0_2 & \left( W_{1,k}-\mathrm{i}W_{2,k}\right) 1_2 & 0_2 \\ 
0_2 & 0_2 & 0_2 & 0_2 \\ 
\left( W_{1,k}+\mathrm{i}W_{2,k}\right) 1_2 & 0_2 & 0_2 & 0_2 \\ 
0_2 & 0_2 & 0_2 & 0_2
\end{array}
\right] \mbox{.} 
\]
Here: $A_k$, $Z_k$, $W_k$are real functions; $g_1$ and $g_2$ are real
positive constants, 
\[
e:=\frac{g_1g_2}{\sqrt{g_1^2+g_2^2}}\mbox{.} 
\]

Fields $W_k$ obey to the following equation \cite[p.127]{PTGT}: 
\begin{equation}
\left( -\frac 1{\mathrm{c}}\partial _t^2+\sum\limits_{s=1}^3\partial
_s^2\right) W_{k,\mu }=g_2^2\left( \widetilde{W}_0^2-\sum\limits_{s=1}^3%
\widetilde{W}_s^2\right) +\Lambda \mbox{.}  \label{WW}
\end{equation}

Here 
\[
\widetilde{W}_\nu :=\left[ 
\begin{array}{c}
W_{0,\nu } \\ 
W_{1,\nu } \\ 
W_{2,\nu }
\end{array}
\right] 
\]
with 
\[
W_{0,\nu }:=Z_\nu \frac{g_2}{\sqrt{g_1^2+g_2^2}}+A_\nu \frac{g_1}{\sqrt{%
g_1^2+g_2^2}} 
\]
and $\Lambda $ is the action of other components of field $W$ on $W_{k,\mu }$%
.

Equation (\ref{WW}) looks like the Klein-Gordon equation of field $W_{k,\mu
} $ with mass 
\[
m:=\frac{\mathrm{h}}{\mathrm{c}}g_2\sqrt{\widetilde{W}_0^2-\sum\limits_{\nu
=1}^3\widetilde{W}_\nu ^2} 
\]
and with additional terms of the $W_{k,\mu }$ interactions with other
components of $W$.

This ''mass'' is invariant under Lorentz transformations: 
\begin{eqnarray*}
\widetilde{W}_0 &\rightarrow &\widetilde{W}_0^{\prime }:=\frac{\widetilde{W}%
_0-\frac v{\mathrm{c}}\widetilde{W}_k}{\sqrt{1-\left( \frac v{\mathrm{c}%
}\right) ^2}}\mbox{, }\widetilde{W}_k\rightarrow \widetilde{W}_k^{\prime }:=%
\frac{\widetilde{W}_k-\frac v{\mathrm{c}}\widetilde{W}_0}{\sqrt{1-\left(
\frac v{\mathrm{c}}\right) ^2}}\mbox{,} \\
\widetilde{W}_j &\rightarrow &\widetilde{W}_j^{\prime }:=\widetilde{W}_j%
\mbox{, if }j\neq k\mbox{,}
\end{eqnarray*}
is invariant under turns of the $\left( \widetilde{W}_1,\widetilde{W}_2,%
\widetilde{W}_3\right) $ space: 
\[
\left\{ 
\begin{array}{c}
\widetilde{W}_k\rightarrow \widetilde{W}_k^{\prime }:=\widetilde{W}_k\cos
\lambda -\widetilde{W}_s\sin \lambda \mbox{,} \\ 
\widetilde{W}_s\rightarrow \widetilde{W}_s^{\prime }:=\widetilde{W}_s\cos
\lambda +\widetilde{W}_k\sin \lambda \mbox{,}
\end{array}
\right| 
\]
and invariant under a global weak isospin transformation $U^{(-)}$ 
\cite[p.30]{NH}: 
\[
W_\nu \rightarrow W_\nu ^{\prime }:=U^{(-)}W_\nu U^{(-)\dagger } 
\]
but is not invariant for a local transformation $U^{(-)}$. But local
transformations for $W_{0,\mu }$, $W_{1,\mu }$, $W_{2,\mu }$ are
insignificant since all three particles are very short-lived and a
measurement of masses of these particles is practically possible only at the
point $\left( t\approx 0,\mathbf{x}\approx 0\right) $.

If 
\[
\alpha :=\arctan \frac{g_1}{g_1} 
\]
then masses of $Z$ and $W$ fulfill the following ratio: 
\[
m_W=\frac{m_Z}{\cos \alpha }\mbox{.} 
\]

Therefore, the Glashow's electroweak theory without Higgs is deduced from
properties of physics events probabilities.

\section{Chromatic states and gluons}

The following part of (\ref{ham0}) I call \textit{chromatic movement equation%
}:

\begin{equation}
\left( 
\begin{array}{c}
\sum_{k=0}^3\beta ^{\left[ k\right] }\left( -\mathrm{i}\partial _k+\Theta
_k+\Upsilon _k\gamma ^{\left[ 5\right] }\right) - \\[2pt] 
-M_{\zeta ,0}\gamma _\zeta ^{[0]}+M_{\zeta ,4}\zeta ^{[4]}\,+ \\[2pt] 
-M_{\eta ,0}\gamma _\eta ^{[0]}-M_{\eta ,4}\eta ^{[4]}\,+ \\[2pt] 
+M_{\theta ,0}\gamma _\theta ^{[0]}+M_{\theta ,4}\theta ^{[4]}
\end{array}
\right) \varphi =0\mbox{.}  \label{clrH}
\end{equation}

Here \cite{Q1}: 
\[
\gamma _\zeta ^{[0]}=-\left[ 
\begin{array}{cc}
0_2 & \sigma _1 \\ 
\sigma _1 & 0_2
\end{array}
\right] \mbox{, }\zeta ^{[4]}=\mathrm{i}\left[ 
\begin{array}{cc}
0_2 & \sigma _1 \\ 
-\sigma _1 & 0_2
\end{array}
\right] 
\]
are mass elements of red pentad;

\[
\gamma _\eta ^{[0]}=-\left[ 
\begin{array}{cc}
0_2 & \sigma _2 \\ 
\sigma _2 & 0_2
\end{array}
\right] ,\eta ^{[4]}=\mathrm{i}\left[ 
\begin{array}{cc}
0_2 & \sigma _2 \\ 
-\sigma _2 & 0_2
\end{array}
\right] 
\]
are mass elements of green pentad;

\[
\gamma _\theta ^{[0]}=-\left[ 
\begin{array}{cc}
0_2 & \sigma _3 \\ 
\sigma _3 & 0_2
\end{array}
\right] \mbox{, }\zeta ^{[4]}=\mathrm{i}\left[ 
\begin{array}{cc}
0_2 & \sigma _3 \\ 
-\sigma _3 & 0_2
\end{array}
\right] 
\]
are mass elements of blue pentad;

I call:

\begin{itemize}
\item  $M_{\zeta ,0}$, $M_{\zeta ,4}$ \textit{red lower and upper mass
members};

\item  $M_{\eta ,0}$, $M_{\eta ,4}$ \textit{green lower and upper mass
members};

\item  $M_{\theta ,0}$, $M_{\theta ,4}$ \textit{blue lower and upper mass
members}.
\end{itemize}

The mass members of this equation form the following matrix sum: %
\vspace*{-2pt} 
\[
\widehat{M}:= \left( 
\begin{array}{c}
-\,M_{\zeta ,0}\gamma _\zeta ^{[0]}+M_{\zeta ,4}\zeta ^{[4]}\,- \\[3pt] 
-\,M_{\eta ,0}\gamma _\eta ^{[0]}-M_{\eta ,4}\eta ^{[4]}\,+ \\[3pt] 
+\,M_{\theta ,0}\gamma _\theta ^{[0]}+M_{\theta ,4}\theta ^{[4]}
\end{array}
\right) = 
\]
\begin{eqnarray*}
&&\rule{-1cm}{0pt}=\ \left[ 
\begin{array}{cccc}
0 & 0 & -M_{\theta ,0} & M_{\zeta ,\eta ,0} \\[3pt] 
0 & 0 & M_{\zeta ,\eta ,0}^{*} & M_{\theta ,0} \\[3pt] 
-M_{\theta ,0} & M_{\zeta ,\eta ,0} & 0 & 0 \\[3pt] 
M_{\zeta ,\eta ,0}^{*} & M_{\theta ,0} & 0 & 0
\end{array}
\right] + \\[4pt]
&&\rule{-1cm}{0pt}+\mathrm{i}\left[ 
\begin{array}{cccc}
0 & 0 & M_{\theta ,4} & M_{\zeta ,\eta ,4}^{*} \\[3pt] 
0 & 0 & M_{\zeta ,\eta ,4} & -M_{\theta ,4} \\[3pt] 
-M_{\theta ,4} & -M_{\zeta ,\eta ,4}^{*} & 0 & 0 \\[3pt] 
-M_{\zeta ,\eta ,4} & M_{\theta ,4} & 0 & 0
\end{array}
\right]
\end{eqnarray*}

with $M_{\zeta ,\eta ,0}:=M_{\zeta ,0}-\mathrm{i}M_{\eta ,0}$ and $M_{\zeta
,\eta ,4}:=M_{\zeta ,4}-\mathrm{i}M_{\eta ,4}$.

Elements of these matrices can be rotated by the following octad elements 
\cite{Q1}:\\$\acute U:=\left\{ U_{1,2}\left( \zeta \right) ,U_{1,3}\left(
\vartheta \right) ,U_{2,3}\left( \alpha \right) ,U_{0,1}\left( \sigma
\right) ,U_{0,2}\left( \phi \right) ,U_{0,3}\left( \iota \right) ,\widetilde{%
U}\left( \chi \right) ,\widehat{U}\left( \kappa \right) \right\} $\\where $%
\zeta \left( t,\mathbf{x}\right) $, $\vartheta \left( t,\mathbf{x}\right) $, 
$\alpha \left( t,\mathbf{x}\right) $, $\sigma \left( t,\mathbf{x}\right) $, $%
\phi \left( t,\mathbf{x}\right) $, $\iota \left( t,\mathbf{x}\right) $, $%
\chi \left( t,\mathbf{x}\right) $, $\kappa \left( t,\mathbf{x}\right) $ are
any real functions.

For example, if 
\begin{eqnarray*}
\widehat{M}^{\prime } &:&=U_{2,3}^{-1}\left( \alpha \right) \widehat{M}%
U_{2,3}\left( \alpha \right) :=\! \\
&&-M_{\zeta ,0}^{\prime }\gamma _\zeta ^{[0]}+M_{\zeta ,4}^{\prime }\zeta
^{[4]}- \\
&&-M_{\eta ,0}^{\prime }\gamma _\eta ^{[0]}-M_{\eta ,4}^{\prime }\eta ^{[4]}+
\\
&&+M_{\theta ,0}^{\prime }\gamma _\theta ^{[0]}+M_{\theta ,4}^{\prime
}\theta ^{[4]}\!\!\!
\end{eqnarray*}
then \vspace*{4pt} 
\begin{eqnarray*}
&&\rule{-1cm}{0pt} M_{\zeta ,0}^{\prime }=M_{\zeta ,0}\,, \\
&&\rule{-1cm}{0pt} M_{\eta ,0}^{\prime }=M_{\eta ,0}\cos 2\alpha +M_{\theta
,0}\sin 2\alpha \,, \\
&&\rule{-1cm}{0pt} M_{\theta ,0}^{\prime }=M_{\theta ,0}\cos 2\alpha
-M_{\eta ,0}\sin 2\alpha \,, \\
&&\rule{-1cm}{0pt} M_{\zeta ,4}^{\prime }=M_{\zeta ,4}\,, \\
&&\rule{-1cm}{0pt} M_{\eta ,4}^{\prime }=M_{\eta ,4}\cos 2\alpha +M_{\theta
,4}\sin 2\alpha \,, \\
&&\rule{-1cm}{0pt} M_{\theta ,4}^{\prime }=M_{\theta ,4}\cos 2\alpha
-M_{\eta ,4}\sin 2\alpha \,.
\end{eqnarray*}

Therefore, matrix $U_{2,3}\left( \alpha \right) $ makes an oscillation
between green and blue colors. And this transformation of equation (\ref
{clrH}) bends time-space as the following: 
\begin{eqnarray}
\frac \partial {\partial x_2^{\prime }} &:&=\cos \left( 2\alpha \right)
\frac \partial {\partial x_2}-\sin \left( 2\alpha \right) \frac \partial
{\partial x_3}\mbox{,}  \label{u23} \\
\frac \partial {\partial x_3^{\prime }} &:&=\cos \left( 2\alpha \right)
\frac \partial {\partial x_3}+\sin \left( 2\alpha \right) \frac \partial
{\partial x_2}\mbox{.}  \nonumber
\end{eqnarray}

One more example: if 
\begin{eqnarray*}
\widehat{M}^{\prime \prime } &:&=U_{0,1}^{-1}\left( \sigma \right) \widehat{M%
}U_{0,1}\left( \sigma \right) :=\! \\
&&-M_{\zeta ,0}^{\prime \prime }\gamma _\zeta ^{[0]}+M_{\zeta ,4}^{\prime
\prime }\zeta ^{[4]}- \\
&&-M_{\eta ,0}^{\prime \prime }\gamma _\eta ^{[0]}-M_{\eta ,4}^{\prime
\prime }\eta ^{[4]}+ \\
&&+M_{\theta ,0}^{\prime \prime }\gamma _\theta ^{[0]}+M_{\theta ,4}^{\prime
\prime }\theta ^{[4]}\!\!\!
\end{eqnarray*}

then: \vspace*{-6pt} 
\begin{eqnarray*}
&&\rule{-.5cm}{0pt} M_{\zeta ,0}^{\prime \prime }=M_{\zeta ,0}\,, \\[2pt]
&&\rule{-.5cm}{0pt} M_{\eta ,0}^{\prime \prime }=\left( M_{\eta ,0}\cosh
2\sigma -M_{\theta ,4}\sinh 2\sigma \right)\,, \\[2pt]
&&\rule{-.5cm}{0pt} M_{\theta ,0}^{\prime \prime }=M_{\theta ,0}\cosh
2\sigma +M_{\eta ,4}\sinh 2\sigma\,, \\[2pt]
&&\rule{-.5cm}{0pt} M_{\zeta ,4}^{\prime \prime }=M_{\zeta ,4}\,, \\[2pt]
&&\rule{-.5cm}{0pt} M_{\eta ,4}^{\prime \prime }=M_{\eta ,4}\cosh 2\sigma
+M_{\theta ,0}\sinh 2\sigma\,, \\[2pt]
&&\rule{-.5cm}{0pt} M_{\theta ,4}^{\prime \prime }=M_{\theta ,4}\cosh
2\sigma -M_{\eta ,0}\sinh 2\sigma\,.
\end{eqnarray*}

Therefore, matrix $U_{0,1}\left( \sigma \right) $ makes an oscillation
between green and blue colours with an oscillation between upper and lower
mass members. And this transformation of equation (\ref{clrH}) bends
time-space as the following: 
\begin{equation}
\left. 
\begin{array}{l}
\displaystyle \frac{\partial x_1}{\partial x_1^{\prime }}=\cosh 2\sigma \\%
[10pt] 
\displaystyle \frac{\partial t}{\partial x_1^{\prime }}=\frac 1{\mathrm{c}%
}\sinh 2\sigma \\[10pt] 
\displaystyle \frac{\partial x_1}{\partial t^{\prime }}=\mathrm{c}\sinh
2\sigma \\[10pt] 
\displaystyle \frac{\partial t}{\partial t^{\prime }}=\cosh 2\sigma \\[10pt] 
\displaystyle \frac{\partial x_2}{\partial t^{\prime }}=\frac{\partial x_3}{%
\partial t^{\prime }}=\frac{\partial x_2}{\partial x_1^{\prime }}=\frac{%
\partial x_3}{\partial x_1^{\prime }}=0
\end{array}
\right\} .  \label{grg}
\end{equation}

Therefore, the oscillation between blue and green colors with the
oscillation between upper and lower mass members bends the space in the $t$, 
$x_1$ directions.

Such transformation with elements of set $\acute U$ add to equation (\ref
{clrH}) gauge fields of the following shape: $U_k^{-1}\left( \xi \right)
\partial _sU_k\left( \xi \right) $ where: $U_k\left( \xi \right) \in \acute
U $. And for every element $U_k\left( \xi \right) $ of $\acute U$ exists 
\cite{Q1} matrix $\Lambda _k$ such that 
\[
U_k^{-1}\left( \xi \right) \partial _sU_k\left( \xi \right) =\Lambda
_k\partial _s\xi \mbox{,} 
\]
and for every product $U$ of $\acute U$'s elements real functions $G_s^r(t,%
\mathbf{x})$ exist such that 
\[
U^{-1}\left( \xi \right) \partial _sU\left( \xi \right) =\frac{g_3}%
2\sum\limits_{r=1}^8\Lambda _rG_s^r 
\]
with some real constant $g_3$ (similar to 8 gluons).

\section{Asymptotic freedom, confinement,\protect\\Newton's gravity}

From (\ref{grg}): \vspace*{-6pt} 
\begin{eqnarray*}
&&\rule{-.5cm}{0pt} \frac{\partial x_1}{\partial t^{\prime }} =\mathrm{c}%
\sinh 2\sigma \,, \\[2pt]
&&\rule{-.5cm}{0pt} \frac{\partial t}{\partial t^{\prime }} =\cosh 2\sigma
\,.
\end{eqnarray*}

\vspace*{-2pt} Because \vspace*{-3pt} 
\begin{eqnarray*}
&&\rule{-.5cm}{0pt} \sinh 2\sigma =\frac v{\sqrt{1-\frac{v^2}{\mathrm{c}^2}}%
}\,, \\
&&\rule{-.5cm}{0pt} \cosh 2\sigma =\frac 1{\sqrt{1-\frac{v^2}{\mathrm{c}^2}}}
\end{eqnarray*}
where $v$ is a velocity of system $\left\{ t^{\prime },x_1^{\prime }\right\} 
$ as respects system $\left\{ t,x_1\right\} $ then 
\[
v=\mathrm{c}\tanh 2\sigma \,. 
\]

Let 
\[
2\sigma :=\omega \left( x_1\right) \frac t{x_1} 
\]

\vspace*{-4pt}\noindent
with 
\[
\omega \left( x_1\right):=\frac \lambda {\left| x_1\right| }\,, 
\]
where $\lambda$ is a real constant bearing positive numerical value.

In that case 
\begin{figure}[t]
\centering
\includegraphics[natheight=702px, natwidth=1123px, width=120mm]{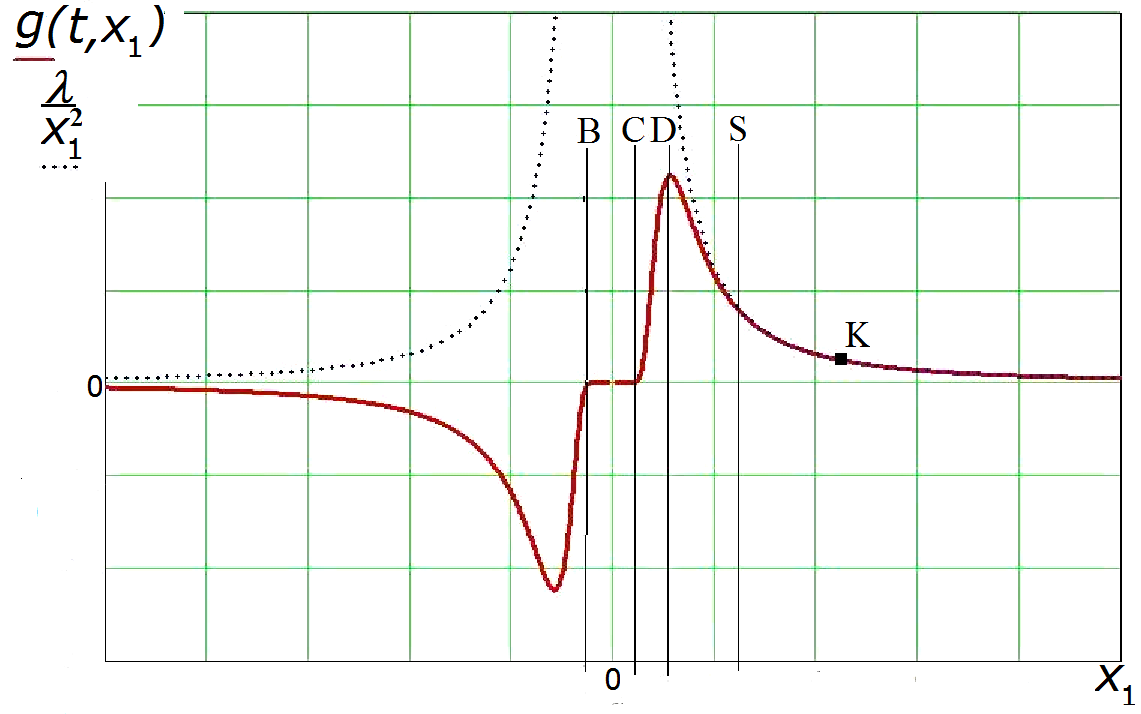}
\par
\vspace*{-3.5mm}
\par
\vspace*{-2mm}
\caption{Dependency of $g(t,x_1)$ from $x_1$.}
\end{figure}

\begin{figure}[t]
\centering
\includegraphics[natheight=702px, natwidth=1123px, width=120mm]{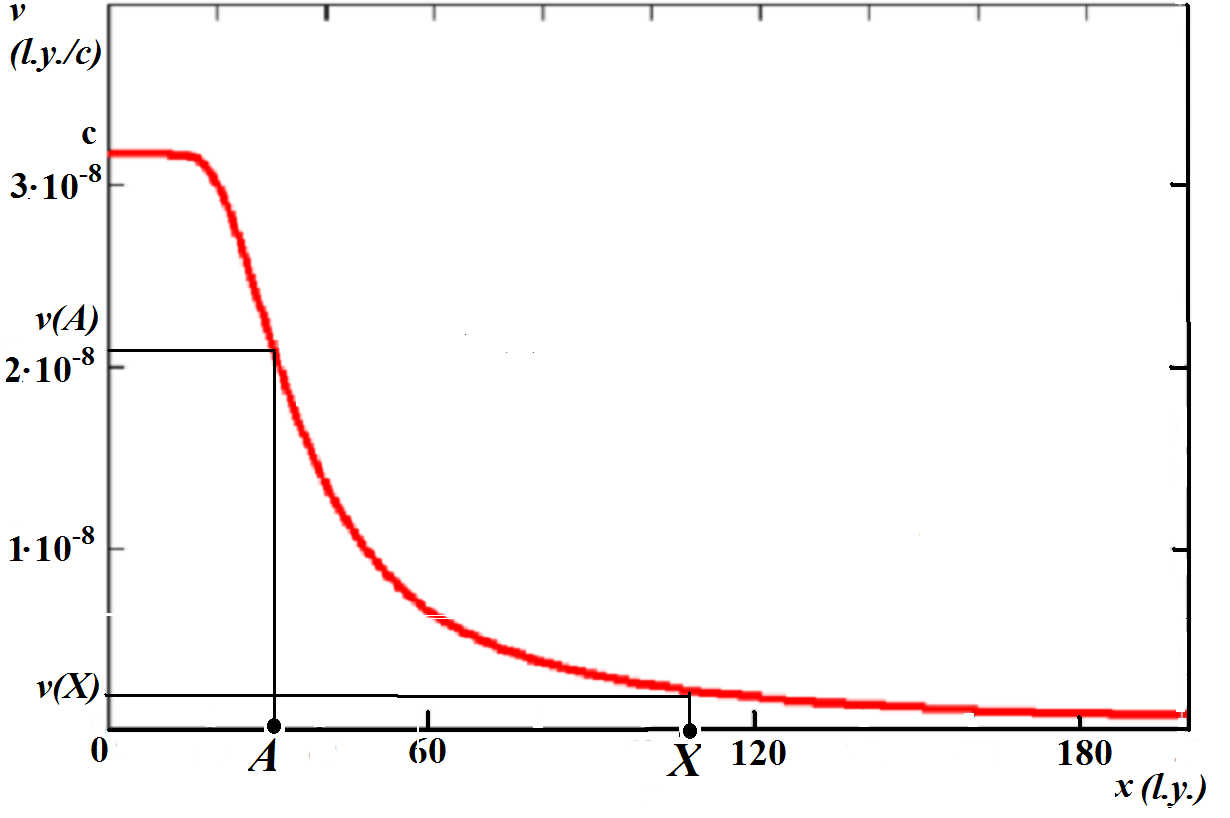}
\par
\vspace*{-3.5mm}
\par
\vspace*{-2mm}
\caption{Dependency of $v(t,x)$ from $x$}
\end{figure}

\begin{figure}[t]
\centering
\includegraphics[natheight=702px, natwidth=1123px, width=120mm]{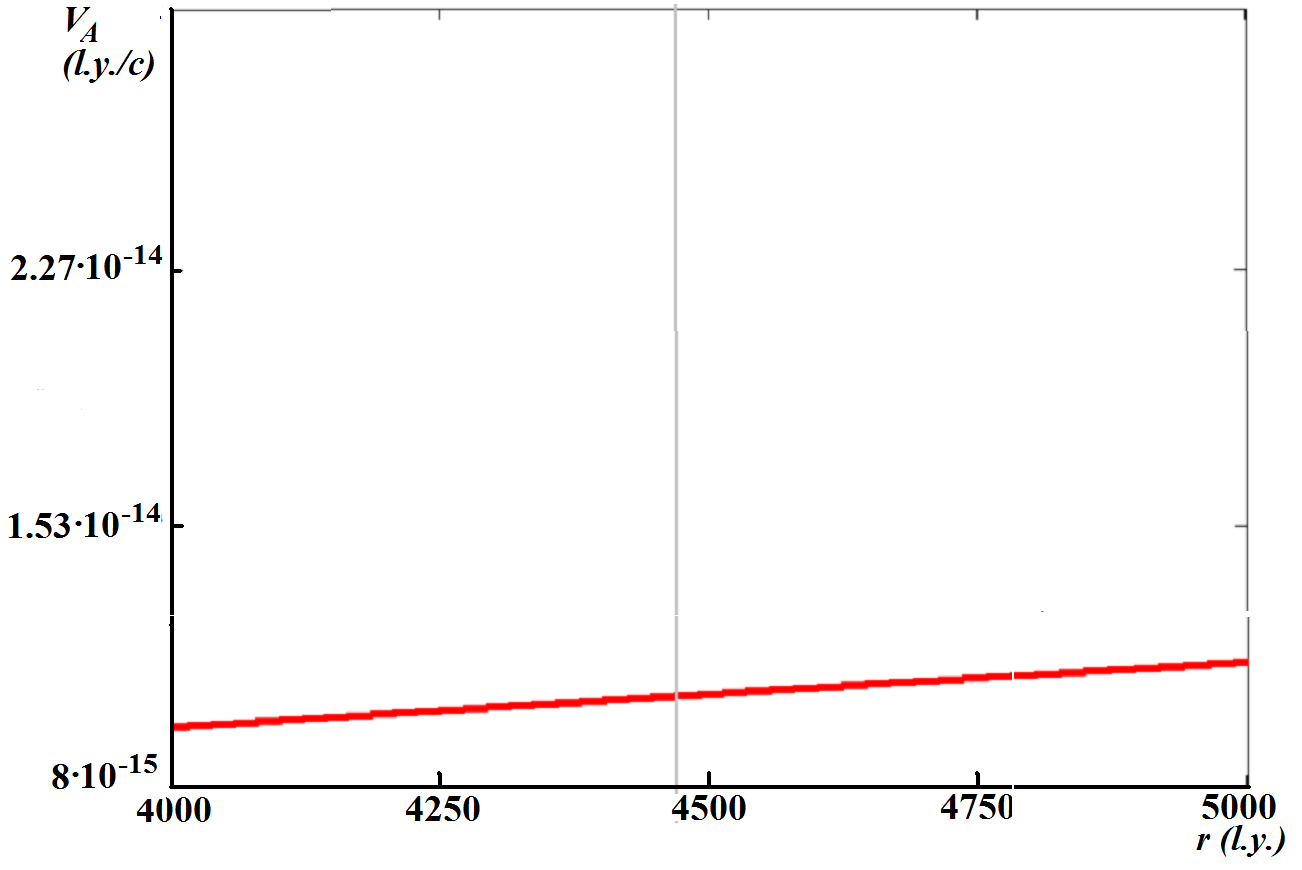}
\par
\vspace*{-3.5mm}
\par
\vspace*{-2mm}
\caption{Dependence of $V_A\left( r\right) $ on $r$ with $x_A=25\times 10^3$
l.y.}
\end{figure}

\begin{figure}[t]
\centering
\includegraphics[natheight=702px, natwidth=1123px, width=120mm]{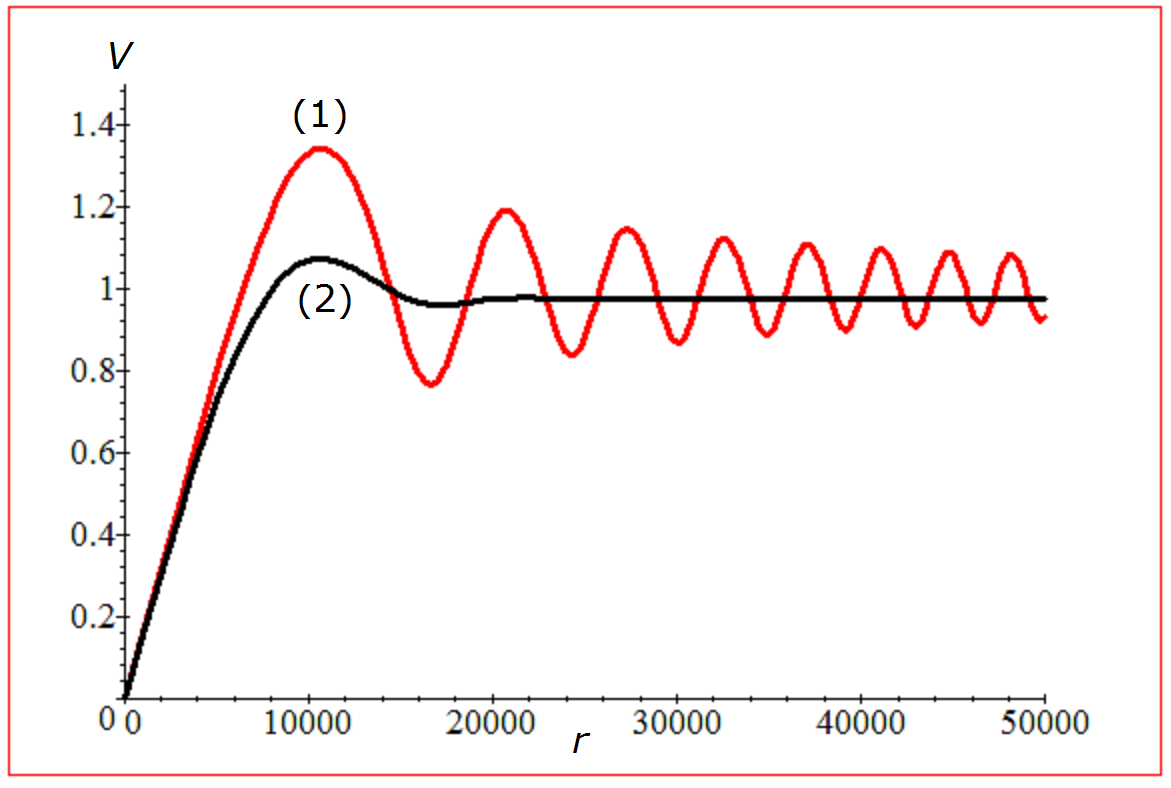}
\par
\vspace*{-3.5mm}
\par
\vspace*{-2mm}
\caption{Dependence of velocity $v$ on the radius $r$ at large $t\sim $~ $%
10^{+8}$ and line (1) at $\theta =\theta $ , and line (2) at $\theta =13\pi
/14$}
\end{figure}

\[
v\left( t,x_1\right) =\tanh \left( \omega \left( x_1\right) \frac
t{x_1}\right) 
\]
and if $g$ is an acceleration of system $\left\{ t^{\prime },x_1^{\prime
}\right\} $ as respects system $\left\{ t,x_1\right\} $ then \vspace*{-2pt} 
\[
g\left( t,x_1\right) =\frac{\partial v}{\partial t}=\frac{\omega \left(
x_1\right) }{x_1\cosh ^2\left( \omega \left( x_1\right) \frac t{x_1}\right) }%
\,\mbox{.} 
\]

Figure 1 shows a dependency of this acceleration on $x_1$.

If an object immovable in system $\left\langle t,x_1\right\rangle $ is
placed in point $K$ then in system $\left\langle t^{\prime },x_1^{\prime
}\right\rangle $ this object must move to the left with acceleration $g$ and 
\[
g\left( x_1\right) \approx \frac \lambda {x_1^2}\mbox{.} 
\]

I call:

\begin{itemize}
\item  interval from $S$ to $\infty $: \textit{Newton Gravity Zone},

\item  interval from $B$ to $C$: \textit{Asymptotic Freedom Zone},

\item  and interval from $C$ to $D$: \textit{Confinement Force Zone}.
\end{itemize}

if $t$ has some fixed volume, $x>0$, and $\Lambda :=\lambda t$ then 
\begin{equation}
v\left( x\right) =\mathrm{c}\tanh \left( \frac \Lambda {x^2}\right)\mbox{.}
\label{4}
\end{equation}
A dependency of $v(x)$ (light years/c) from $x$ (light years) with $\Lambda
= 741.907$ is shown in Figure 2.

\section{Dark energy and dark matter}

Let a placed in a point $A$ observer be stationary in the coordinate system $%
\left\{ t,x\right\} $. Hence, in the coordinate system $\left\{ t^{\prime
},x^{\prime }\right\} $ this observer is flying to the left to the point $O$
with velocity $-v\left( x_A\right) $. And point $X$ is flying to the left to
the point $O$ with velocity $-v\left( x\right) $.

Consequently, the observer $A$ sees that the point $X$ flies away from him
to the right with velocity 
\begin{equation}
V_A\left( x\right) =\mathrm{c}\tanh \left( \frac \Lambda {x_A^2}-\frac
\Lambda {x^2}\right)  \label{6}
\end{equation}
in accordance with the relativistic rule of addition of velocities.

Let $r:=x-x_A$ (i.e. $r$ is distance from $A$ to $X$), and 
\begin{equation}
V_A\left( r\right) :=\mathrm{c}\tanh \left( \frac \Lambda {x_A^2}-\frac
\Lambda {\left( x_A+r\right) ^2}\right)\mbox{.}  \label{7}
\end{equation}

In that case Figure 3 demonstrates the dependence of $V_A\left( r\right) $
on $r$ with $x_A=25\times 10^3$ l.y.

Hence, $X$ runs from $A$ with almost constant acceleration: 
\begin{equation}
\frac{V_A\left( r\right) }r=H\mbox{.}  \label{8}
\end{equation}

Therefore, the phenomenon of the accelerated expansion of Universe is
explained by oscillations of chromatic states.

From (\ref{u23}): dependence of velocity of rotation of galactic objects on
distance to the galaxy center is given by a Figure 4 \cite{DM}.

\section{Events and particles}

Thus, concepts and statements of Quantum Theory are concepts and statements
of the probability of pointlike events and their ensembles.

Elementary physical particle in vacuum behaves like these probabilities. For
example, accordance to doubleslit experiment \cite{Arx}, if partition with
two slits is placed between a source of elementary particles and a detecting
screen in vacuum then interference occurs. But if this system will be put in
a cloud chamber, then trajectory of particle will be clearly marked with
drops of condensate and any interference will disappear. It seems that a
physical particle exists only in the instants of time when some events
happen to it. And in the other instants of time the particle does not exist,
but the probability of some event to happen to this particle remains.

Thus, if no event occurs between an event of creation of a particle and an
event of detection of it, then the particle does not exist in this period of
time. There exists only the probability of detection of this particle at
some point. But this probability, as we have seen, obeys the equations of
quantum theory and we get the interference. But in a cloud chamber events of
condensation form a chain meaning the trajectory of this particle. In this
case the interference disappears. But this trajectory is not continuous -
each point of this line has an adjacent point. And the effect of movement of
this particle arises from the fact that a wave of probability propagates
between these points.

Consequently, the elementary physical particle represents an ensemble of
pointlike events associated with probabilities. And charge, mass, energy,
momentum, spins, etc. represent parameters of distribution of these
probabilities. It explains all paradoxes of quantum physics. Schrodinger's
cat lives easily without any superposition of states until the microevent
awaited by everyone occures. And the wave function disappears without any
collapse in the moment when event probability disappears after the event
occurs.

Hence, entanglement concerns not particles but probabilities. That is when
event of the measuring of spin of Alice's electron occurs then probability
for these entangled electrons is changed instantly on whole space.
Therefore, nonlocality acts for probabilities, not for particles. But
probabilities can not transmit any information.


Thus, the fundamental entities of nature are not particles and fields, but
pointlike events and probability connecting these events.

\section{Conclusion}

All known concepts and statements of fundamental physics can be obtained
from the properties of the probability of pointlike events. Thus any new
particles or forces it is impossible.

\end{document}